\documentstyle[prl,aps,multicol,epsf]{revtex}
\begin{document}

\title{A phenomenology for
multiple phases in the heavy fermion
skutterudite superconductor PrOs$_4$Sb$_{12}$}
\author{Jun Goryo}
\address{Max Planck Institute for the Physics of Complex Systems,
N{\"o}thnitzer Str. 38, 01187 Dresden, Germany}

%{\today}
\maketitle
\begin{abstract}
The two superconducting phases recently discovered
in skutterudite PrOs$_4$Sb$_{12}$ are discussed by using
the phenomenological Ginzburg-Landau theory based on the  
the cubic (tetrahedral) crystal symmetry T$_h$. Building on  experimental input 
coming from the
recent thermal conductivity measurements, the structure of the
gap functions for these two phases are considered for the spin singlet
pairing case. One of
the phases, lying in the high temperature and high field region (A-phase)  
is a strongly "anisotropic $s$-wave" state, whereas the other  
phase (B-phase) is an "anisotropic $s+ i d$-wave" state.
The B-phase breaks cubic crystal symmetry and time reversal symmetry
spontaneously. The cubic symmetry breaking in B-phase causes
anomalous crystal strains which could be seen in the anisotropy of
thermal expansions and isothermal elastic constants. 
As a signal of the 
time reversal symmetry breaking, it is expected that an internal spontaneous 
magnetization should be observed by the $\mu$SR measurement.

\end{abstract}


\begin{multicols}{2}

\section{Introduction}

\begin{figure}
\centerline{
\epsfysize=6cm\epsffile{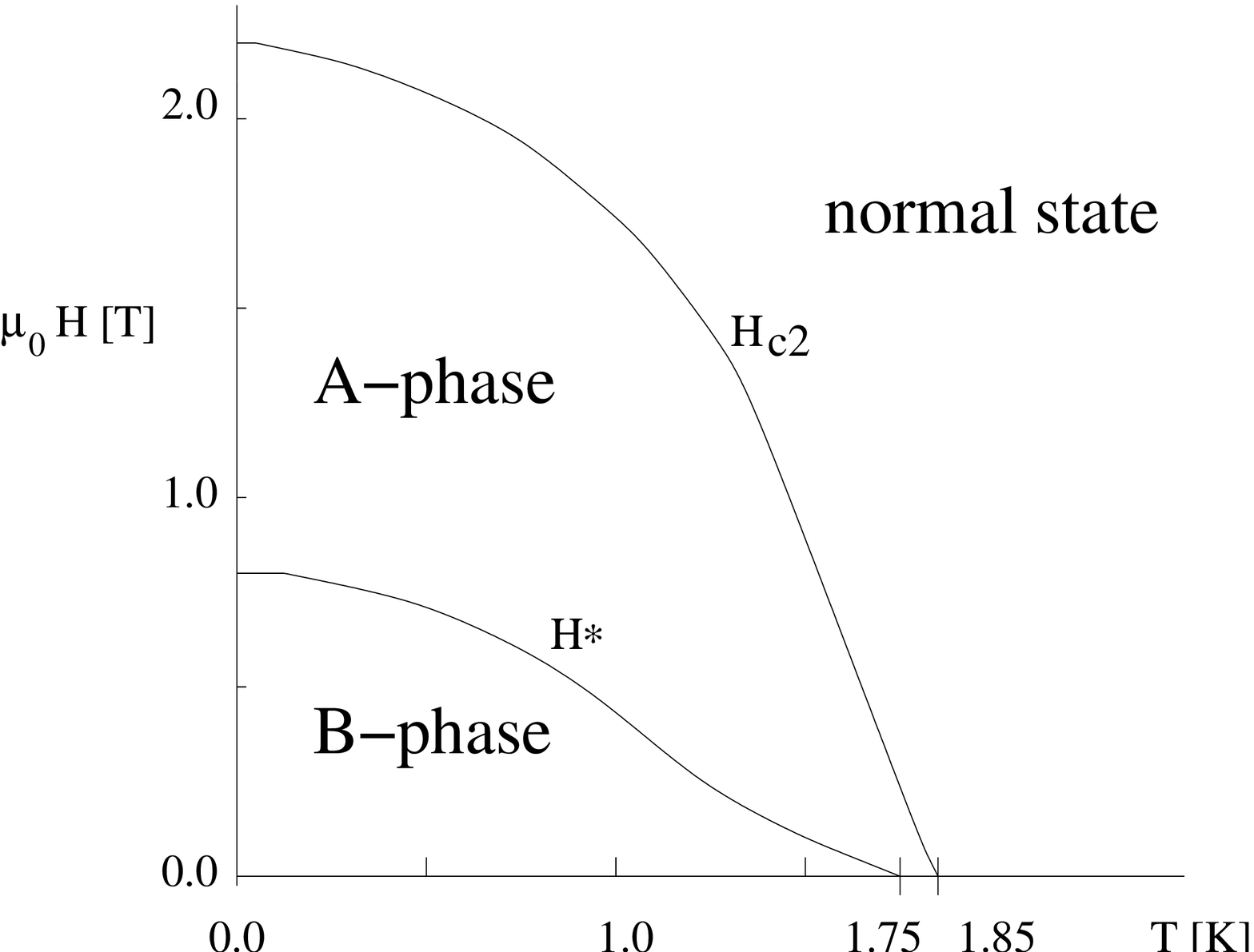}}
\vspace{1cm}
FIG. I. A rough sketch of the phase diagram
proposed by Izawa et. al\cite{Izawa}. See, also Refs. \cite{Vollmer} and
\cite{Niels}.
\label{phase-diagram}
\end{figure}

Recently, heavy fermion superconductivity
was discovered at 1.85 $K$
in the skutterudite compound PrOs$_4$Sb$_{12}$ with a cubic
(tetrahedral) crystal symmetry T$_h$\cite{Bauer}.
Intriguingly, it was indicated
that there are multiple superconducting phases
in this material, like in the Uranium compounds\cite{Sigrist-Ueda,Joynt}.
Two specific heat jumps were 
observed at $T_{c1}=1.85K$ and $T_{c2}=1.75K$
in the absence of the magnetic field\cite{Vollmer}.
Moreover, the thermal conductivity measurement in a
magnetic field rotated relative to the crystal axes reveals
a novel change in the symmetry of the gap function
in the $ab$-plane\cite{Izawa}.
The result was that the gap function has
four point-node-like structure (four dips) in the [100] and [010]
directions on the Fermi surface
in the high field and high temperature region (A-phase), and
two dips in the [010] direction in the low field and
low temperature region (B-phase).
The proposed phase diagram is given in Fig. \ref{phase-diagram}.
Unfortunately, the gap structure
along the $c$-axis remains unresolved within in
the presently available  data\cite{Izawa}, and one of the purpose of
this paper is to consider this problem.
The presence of the point node like structures appears
to be consistent with the power law behavior of
the specific heat\cite{Bauer,Vollmer} and
the nuclear-spin-lattice relaxation\cite{Kotegawa}
at low temperatures (See also Ref. \cite{Maclaughlin}).
Another evidence for the double superconducting transition is
the thermal expansion measurement along [100] direction\cite{Niels}; 
 two clear jumps are observed at 
around $T_{c1}=1.85K$ and $T_{c2}=1.75K$
in the absence of the magnetic field (Fig. \ref{phase-diagram}).

In this paper, the symmetry of the gap functions
for A- and B-phase are discussed by use of a  
phenomenological Ginzburg-Landau (GL) theory\cite{Sigrist-Ueda} 
based on the tetrahedral crystal symmetry T$_h$.
At  present, there are no information
as to whether the superconductivity is spin singlet or spin triplet
pairing. Only the singlet pairing case will be discussed in this paper.
The gap structure along the $c$-axis is considered.
We find that the gap function appears to have two additional dips
along the $c$-axis in A-phase 
(in total there are six dips, along the [100], [010]
and [001] directions) ,
while no additional dips are inferred for the  
B-phase resulting in a toal of two dips, along the [010] direction) .
The A-phase gap is essentially an "anisotropic $s$-wave" state
and the B-phase gap is an "anisotropic $s + i d$-wave" state.
The B-phase gap breaks the cubic crystal symmetry and time reversal symmetry
spontaneously,  from which one can expect a crystal deformation
with cubic symmetry breaking to occur in this phase. 
To verify this, the phenomenological
strain-order parameter coupling is taken into account and
the thermal expansions and isothermal elastic constants
are calculated\cite{Thalm}.
One finds that these values become anisotropic in
B-phase.

%%%%%%%%%%%%%%%%%%%%%%%%%%%%%%%%%%%%%%%%%%%%%%%%%%%%%%%%%%%%%%%%%%%%%%%%
%\vspace{0cm}
%%%%%%%%%%%%%%%%%%%%%%%%%%%%%%%%%%%%%%%%%%%%%%%%%%%%%%%%%%%%%%%%%%%%%%%%%

The pairing mechanism for this superconductivity should be also interesting.
It is widely believed that
the mechanism for the heavy fermion superconductivity in
the Uranium compounds comes from magnetic interactions, but
PrOs$_4$Sb$_{12}$ is a non-magnetic compound.
Several experiments indicate that
the ground state of
Pr $4f$-electrons in the crystalline electric field
would most likely be the non-Kramers doublet $\Gamma_3$
state\cite{Bauer,Vollmer,Maclaughlin}.
The $\Gamma_3$ ground state has quadrupolar fluctuations and
its relations to the heavy fermion behavior\cite{Cox} and also
to a pairing mechanism seems to be possible to discuss.
However, its relation to the pairing symmetry, which will be focused in this
paper, is unclear at present.
Moreover, the singlet $\Gamma_1$ state without quadrupolar fluctuations
is also proposed as the ground state\cite{Aoki}. The details of
the pairing mechanism will not be discussed here.

The organization of this paper is as follows.
In section II, the gap structures for the A- and B-phases 
are discussed. In section III, jumps of the specific heat
at two transition points (T$_{c1}$ and T$_{c2}$) are calculated.
To fit these calculations to the specific heat measurement\cite{Vollmer},
one obtains information for the coefficients of the fourth order
terms in the GL potential. In section IV, arguments for a spontaneous
cubic symmetry breaking in the B-phase are presented, and a 
summary is given in section V.

\section{The Gap structures}

In this section, the gap structures for the 
A- and B-phases in the spin singlet pairing case are discussed
within the frameork of the phenomenological GL theory.
Information on the gap structure along the $c$-axis, which
is still unresolved by the thermal conductivity
measurement\cite{Izawa}, are obtained.

The GL theory to be considered is based on 
the cubic crystal symmetry group T$_h$.
This group has essentially three irreducible representations
A$_1$ ($\Gamma_1$), E ($\Gamma_2, \Gamma_3$) and T ($\Gamma_5$).

\subsection{A-phase}

In this phase, four point-node-like (four-dip) structures of the
gap function located within  the $ab$-plane are observed along the [100] and [010] directions \cite{Izawa}. 
Then, the question is whether the gap function
has two additional dips along $c$-axis (i.e. the gap has
a total of six dips) or not (i.e.  four dips in all).

It should be pointed out that H$_{c2}$ is isotropic in this
superconductor\cite{Izawa}.
In general, the superconducting states represented by a 
multicomponent order parameter should have an anisotropic H$_{c2}$
\cite{Sigrist-Ueda,Joynt}.
Therefore, the gap function in the A-phase appears to be represented by
a single component order parameter, i.e., the A$_1$ representation.

The gap function with six dips along the [100], [010] and [001] directions
belongs to the A$_1$ representation because this structure has the cubic symmetry
T$_h$. Actually, one can see this as follows.
Let $\eta$ and $\phi_s({\bf k})$ stand for the order parameter
and the basis function of the
A$_1$ representation, respectively.
By the definition, $\phi_s({\bf k})$ should be invariant under T$_h$.
The gap function in this representation is
$$
\Delta({\bf k})=\eta \phi_s({\bf k}).
$$
The six-dip state can be represented by using invariants\cite{Maki}: 
\begin{equation}
\phi_s({\bf k})=s + k_x^4 + k_y^4 + k_z^4 ~~(s\simeq -1),
\label{6p1}
\end{equation}
and also
\begin{equation}
\phi_s({\bf k})=s + k_x^2 k_y^2 + k_y^2 k_z^2 + k_z^2 k_x^2 ~~(s<<1),
\label{6p2}
\end{equation}
$\cdot\cdot\cdot$ etc. are possible.
When one takes $s=-1$ in Eq. (\ref{6p1}) ($s=0$ in Eq. (\ref{6p2})),
the dips become point nodes. Gap functions such as in Eqs. (\ref{6p1}) 
and (\ref{6p2}) have been discussed 
in the context of the YNiB$_2$C$_2$ superconductor\cite{Borocarbides}. 
{\it It should be noted that the function $\phi_s({\bf k})$ 
which can give rise to six dips is not limited to these two forms.} 
There are many possible forms which have six dips. 
The point is that the six-dip state can be described in the A$_1$ 
representation and essentially categorized into the so-called 
"anisotropic $s$-wave" state.

The four-dip state does not belong to the A$_1$ representation
because this gap structure does not have the cubic symmetry.
Actually, one would see in the next subsection that
this state belongs to the A$_1 \oplus$E combined representation\cite{Maki}.
Then, a multicomponent orderparameter is needed to
describe this gap function (See, Eq. (\ref{A+E})), and in general, an 
anisotropy in $H_{c2}$ is to be expected\cite{Sigrist-Ueda,Joynt}.
in contradiction to the experimental result\cite{Izawa}.
Moreover, in the next subsection one would see that
the four-dip structure arises in a somewhat accidental way 
(See, Eq. (\ref{4P})).

Then, in the A-phase, it seems to be natural to consider that
the six-dip state (strongly anisotropic $s$-wave state) is
realized.

\subsection{B-phase}

In this phase, two dips of the 
gap function within in $ab$-plane along [010] direction are
observed\cite{Izawa}. Then, the question is whether the gap function
has two additional dips along the $c$-axis (i.e. the gap has
in total four dips) or not (i.e. a total of two dips).

Phase transitions are observed by varying $H$ and
also varying $T$ without a magnetic field turned on (Fig. \ref{phase-diagram}).
To describe the phase transition without external fields,
one can consider the combined representations.
Possible combinations in the present case are
A$_1 \oplus$E, A$_1 \oplus$ T, E$\oplus$T, and A$_1 \oplus$ E $\oplus$ T.

As we will see below, both the four-dip and two-dip
states belong to the A$_1 \oplus$ E representation.
Let $g_{\rm A_1}$ and $g_{\rm E}$ stand for
the effective coupling constant of the pairing interaction
for the A$_1$ channel and the E channel,
respectively. In general, these values are split by the crystal
field. The A$_1 \oplus$ E combined state is realized
when\cite{Sigrist-Ueda,Sigrist-Rice-89}
\begin{equation}
g_{\rm A_1} \simeq g_{\rm E},
\label{coupling}
\end{equation}
is satisfied. This comes from some accidental degeneracy.

The gap function in A$_1 \oplus$ E model is expanded by the
basis functions of A$_1$ and E representations:
\begin{equation}
\Delta({\bf k})=\eta \phi_s({\bf k})
+ \eta_1 \phi_{2y^2 - z^2 - x^2}({\bf k})
+ \eta_2 \phi_{z^2 - x^2}({\bf k}).
\label{A+E}
\end{equation}
The function $\phi_{2y^2 - z^2 - x^2} ({\bf k})$ is the first component
of the 2D basis of the E representation which transforms
like the function $2k_y^2 - k_z^2 - k_x^2 $ under the T$_h$ group.
It has two line nodes parallel to the $ac$-plane.
The function $\phi_{z^2 - x^2} ({\bf k})$ is the second component
of the 2D basis of the E representation which transforms like 
the function $k_z^2 - k_x^2$ under the T$_h$ group.
It has two line nodes crossing at
two points along the [010] direction. Note that  
these functions may include terms beyond quadratic order 
in the momentum ${\bf k}$. 
However the nodal structure of these basis functions remain unaltered, in general,  
even if the higher order terms are included.

To obtain a four-dip state along the [010] and [001]
directions, one must fine-tune the three basis functions into the particular forms
$\phi_s({\bf k})=1 - 2 (k_x^4 + k_y^4 + k_z^4) / 3$,
$\phi_{2 y^2 - z^2 - x^2}({\bf k})=2 k_y^4 - k_z^4 - k_x^4$,
$\phi_{z^2 - x^2}({\bf k})=\sqrt{3} (k_z^4 - k_x^4)$, where the 
second order terms of ${\bf k}$ are absent, 
and also the three order parameters are fixed to take the particular values \cite{Maki}
\begin{equation}  
(\eta, \eta_1,\eta_2) \simeq (1,-\frac{1}{6}, \frac{1}{2 \sqrt{3}}).
\label{4P}
\end{equation}
This looks highly accidental. 
The gap function with four dips along [100] and [010] directions
discussed in the previous subsection can also be obtained
in the same manner\cite{Maki}.

Two-dip state is realized when
\begin{equation} 
(\eta, \eta_1,\eta_2)=(\alpha, 0, \pm i \beta), ~\alpha, \beta \in {\bf R}, 
\label{2P}
\end{equation}
because $\phi_{z^2 - x^2}({\bf k})$ is zero and $\phi_s({\bf k})$ becomes 
minimum along [010] direction.
Note that the two-dip structure appears generically for all values of the  
two parameters $\alpha$ and $\beta$.
Moreover, it is not necessary to fix
the basal functions $\phi_s({\bf k})$, $\phi_{2y^2 - z^2 - x^2}({\bf k})$
and $\phi_{z^2 - x^2}({\bf k})$ to particular forms.
The only assumption needed here is the six-dip structure of
the function $\phi_s({\bf k})$. 
It should be noted again that the function $\phi_s({\bf k})$ 
is not limited to a particular form to obtain six dips. 
There are many possible forms to obtain six dips and 
described in the A$_1$ representation (anisotropic $s$-wave state).  
Eq. (\ref{6p1}) (or Eq. (\ref{6p2})) is one of the examples. 
Then, the state Eq. (\ref{2P}) is an 
anisotropic $s + i d_{z^2 - x^2}$-wave state
and breaks the cubic crystal symmetry and also
the time reversal symmetry spontaneously.

Let us discuss the stability of these
two states Eqs. $(\ref{4P})$ and $(\ref{2P})$. 
It will be shown that the state Eq. $(\ref{2P})$ is found as 
a stable state of the system, but the state Eq. $(\ref{4P})$ would be not.  

Consider the GL potential for A$_1 \oplus$ E representation of T$_h$
\cite{Sigrist-Rice-89}
\begin{eqnarray}
F_{\rm op}&=&A_{\rm A_1}(T)|\eta|^2 + \beta |\eta|^4 +
A_{\rm E}(T) (|\eta_1|^2 + |\eta_2|^2)
\nonumber\\
&&+ \beta_1 (|\eta_1|^2 + |\eta_2|^2)^2
+\beta_2 (\eta_1 \eta_2^* - \eta_1^* \eta_2)^2
\nonumber\\
&&+ \theta_1 |\eta|^2 (|\eta_1|^2 + |\eta_2|^2)
\nonumber\\
&&+
\frac{\theta_2}{2} \{\eta^{*2} (\eta_1^2 + \eta_2^2) + c.c\},
\label{GLop}
\end{eqnarray}
where
$$
A_{\rm A_1}(T)=\frac{T}{T_{\rm A_1}} - 1,
~A_{\rm E}(T)=\frac{T}{T_{\rm E}} -1,
$$
and
$$
T_{\rm A_1} \sim \exp[- 1 / g_{\rm A_1} N(0)], ~
T_{\rm E}\sim \exp[- 1 / g_{\rm E} N(0)], 
$$
in the weak coupling approximation  (See. Eq. (\ref{coupling})).
$N(0)$ is the density of states
at the Fermi surface.  We consider
a temperature region $T< min[T_{\rm A_1}, T_{\rm
E}]$. The terms proportional to $\theta_1$ and $\theta_2$ denote
the coupling between the order paremeter in the A$_1$ representation
and the two order parameters in the E representation. In general, another
coupling term
$$
\theta_3 \left[e^{i \gamma_3} \eta^* (\eta_1 |\eta_1|^2
- 2 \eta_1 |\eta_2|^2 - \eta_1^* \eta_2^2) + c.c. \right]  
$$
could be included in the potential but this term makes the phase
transition 1st order\cite{Sigrist-Rice-89}, which appears to be
inconsistent with the specific heat experiments\cite{Bauer,Vollmer}.
So, this term is excluded here. For simplicity, we take 
\begin{equation}
\theta_1 = \theta_2.
\label{8}
\end{equation}
By using the parameterization
$$
\eta=|\eta|, ~\eta_1 = |\tilde{\eta}| \cos \psi e^{i \phi_1},
~\eta_2=|\tilde{\eta}| \sin \psi e^{i \phi_2},
$$
Eq. (\ref{GLop}) becomes
\begin{eqnarray}
F_{op}&=&A_{\rm A_1}(T)|\eta|^2 + A_{\rm E}(T) |\tilde{\eta}|^2
+ \beta |\eta|^4
+ \beta_1 |\tilde{\eta}|^4
\nonumber\\
&&-
\beta_2 |\tilde{\eta}|^4 \sin^2 2 \psi \sin^2 (\phi_1 - \phi_2)
+\theta_1 |\eta|^2 |\tilde{\eta}|^2
\nonumber\\
&&+\theta_2 |\eta|^2 |\tilde{\eta}|^2 \left[\cos^2 \psi \cos 2 \phi_1 +
\sin^2 \psi \cos 2 \phi_2 \right].
\nonumber
\end{eqnarray}
After a bit tedious calculation one obtains all of 
the stable solutions\cite{Sigrist-Rice-89}:

\begin{enumerate}
\item $0<\beta_2$

\begin{eqnarray}
(\eta, \eta_1, \eta_2)&=&(a, \frac{b}{\sqrt{2}} e^{i \phi},
\pm i \frac{b}{\sqrt{2}} e^{i \phi}),
%\nonumber\\
%&&(a, \pm \frac{b}{\sqrt{2}} e^{i \phi},
%\mp i \frac{b}{\sqrt{2}} e^{i \phi})
\label{beta2+}
\end{eqnarray}
where,
$$
a=\sqrt{-\frac{A_{\rm A_1}(T)}{2 \beta}},
b=\pm \sqrt{-\frac{A_{\rm E}(T)}{2 (\beta_1 - |\beta_2|)}},
$$
and $\phi$ is arbitral here.
%It could be fixed by taking into account the
%6-th order couplings between A$_1$ and E representations
%in the GL potential, but it is beyond the scope of our argument.

\item $\beta_2 < 0$

\begin{equation}
(\eta, \eta_1, \eta_2)=
\left\{
\begin{array}{l}
(c,\pm id \cos \psi ,\pm i d \sin \psi) {\rm ~,for ~0<\theta_2}
\cr
(c,\pm d \cos \psi ,\pm d \sin \psi) {\rm ~,for ~\theta_2<0,}
\end{array}
\right.
\label{beta2-}
\end{equation}
where,
$$
c=\sqrt{-\frac{A_{\rm A_1}(T)}{2 \beta}},
d=\sqrt{-\frac{A_{\rm E}(T)}{2 \beta_1}},
$$
and $\psi$ is arbitral.
The degeneracy of $\psi$ is lifted by taking into account
 the 6-th order terms for $(\eta_1, \eta_2)$ as 
a weak perturbation\cite{Sigrist-Ueda}
\begin{eqnarray}
\gamma_1 (|\eta_1|^2 + |\eta_2|^2)^3
+ \gamma_2 (|\eta_1|^2 + |\eta_2|^2) |\eta_1^2 + \eta_2^2|^3
\nonumber\\
+ \gamma_3 |\eta_1|^2|3 \eta_2^2 - \eta_1^2|.
\nonumber
\end{eqnarray}
The coupling terms between $\eta$ and $(\eta_1, \eta_2)$ are
neglected here. Then,
\begin{equation}
\psi=
\left\{
\begin{array}{l}
0,\pi~(0<\gamma_3)
\cr
\pm \frac{\pi}{2}~(\gamma_3<0).
\end{array}
\right.
\label{psi}
\end{equation}

\end{enumerate}

From Eqs. (\ref{A+E}), (\ref{2P}), (\ref{beta2-}) and (\ref{psi}),
one can see that the two-dip state 
(anisotropic $s+id_{z^2 - x^2}$-wave state) is
stable when $\beta_2, \gamma_3<0<\theta_2$.
On the other hand, it is hard to see how the four-dip
state Eq. (\ref{4P}) can be stabilized in the present argument.

The gap structures for the other stable solutions
(Eq. (\ref{beta2+}) and Eq. (\ref{beta2-}) for $\theta_2 < 0$) 
are inconsistent with the 
thermal conductivity measurement\cite{Izawa}. 
Furthermore, based on a consideration for the condensation energy 
within the weak coupling approach, it has been pointed out that 
the $s+ i d$-wave state is energetically favored 
in A$_1 \oplus E$ combined representation\cite{Sigrist-T-broken}.

Fron these considerations, we conclude that the two-dip state 
appears to be realized in the B-phase.

\subsection{phase transition and symmetry breaking}

Let us summarize the results in this section
and try to describe the phase transition in
the absence of the magnetic field.

Put the parameters
\begin{equation}
T_{\rm A_1}=T_{c1}=1.85 K,  ~T_{\rm E}=T_{c2}=1.75 K,
\label{Tc}
\end{equation}
and $\beta_2, \gamma_3<0<\theta_2$ in the GL free energy 
Eq. (\ref{GLop}). For simplicity, we take Eq. (\ref{8}). 
Then, the stationary values of
the order parameters in the A-phase ($T_{c2}<T<T_{c1}$) are
\begin{equation}
\eta=\sqrt{-\frac{A_{\rm A_1}(T)}{2 \beta}}, \eta_1=\eta_2=0,
\label{A}
\end{equation}
and in the B phase ($T<T_{c2}$)
\begin{equation}
\eta=\sqrt{-\frac{A_{\rm A_1}(T)}{2 \beta}}, \eta_1=0,
\eta_2=\pm i \sqrt{- \frac{A_{\rm E}(T)}{2 \beta_1}}.
\label{B}
\end{equation}
From Eqs. (\ref{A+E}), (\ref{A}) and (\ref{B}), 
one can see that the gap function in the A-phase has six dips 
along the [100], [010] and [001]
directions (anisotropic $s$-wave state), and
that in the B-phase there are 
two dips along the [010] direction (anisotropic $s+id_{z^2 - x^2}$-wave state).
The result is consistent with
the thermal conductivity measurement\cite{Izawa}.

{\it The gap function in the B-phase breaks the cubic crystal symmetry
and also the time reversal symmetry.}
The consequence of this cubic symmetry breaking
will be discussed in section IV.   
As a signal of the time reversal symmetry breaking, it is expected that 
a spontaneous magnetization be observed by the $\mu$SR measurement
\cite{Sigrist-Ueda}.  

\section{Specific heat jump}

In this section, we calculate the
jumps of the specific heat at the two transition points
$T_{c1}=1.85K$ and $T_{c2}=1.75K$ in Fig. \ref{phase-diagram} indicated by
the specific heat measurement\cite{Vollmer}. To fit the calculation
to the experimental result the ratio of
the coefficients of 4-th order terms in GL potential 
$\beta/\beta_1$ is obtained. The simplification Eq. (\ref{8}) is used. 

Let us introduce
$$
\vec{\eta}=(\eta, \eta_1,\eta_2).
$$
By using the stationary condition for the free energy Eq. (\ref{GLop})
\begin{equation}
0=\frac{\partial F_{op}}{\partial \vec{\eta}},
\label{stat}
\end{equation}
the specific heat is
\begin{eqnarray}
C(T)&=&C_0(T) - T\frac{\partial^2 F_{op}}{\partial T^{2}}
\nonumber\\
&=&C_0(T) - T\left\{\frac{\partial \vec{\eta}}{\partial T} \cdot 
\frac{\partial^2 F}{\partial \vec{\eta} \partial T}
+ c.c.\right\},
\end{eqnarray}
where $C_0(T)$ is the background specific heat.
Let us define $\Delta C(T)=C(T) - C_0(T)$.
Then, in the A-phase,
\begin{equation}
\Delta_{\rm A} C(T)= \frac{T}{2 \beta T_1^2},
\end{equation}
while in the B-phase,
\begin{equation}
\Delta_{\rm B} C(T)= T\left(\frac{1}{2 \beta T_1^2} + \frac{1}{2 \beta_1
T_3^2}\right).
\end{equation}
Therefore, the ratio of the jumps
at the two transition points is
\begin{equation}
\frac{\Delta_{\rm B} C(T_{c2})-\Delta_{\rm A} C(T_{c2})}
{\Delta_{\rm A} C(T_{c1})}=
\frac{\beta T_{c1}}{\beta_1 T_{c2}}
\end{equation}
The experimental result for this quantity is around 1  
\cite{Vollmer}.
Then, from Eq. (\ref{Tc}), one can obtain the ratio
\begin{equation}
\beta / \beta_1 \simeq 0.9.
\label{b/b1}
\end{equation}

\section{Cubic symmetry breaking in B-phase}

the B-phase gap breaks the cubic crystal symmetry $T_h$ spontaneously.
So, one can expect a 
crystal deformation with the cubic symmetry breaking.
In this section, the strain-order parameter coupling is taken into account and
thermal expansions and isothermal elastic constants
are calculated\cite{Thalm}.
In the B-phase, one obtains anisotropic results for
these values. This comes from the spontaneous crystal symmetry breaking.

\subsection{strain-orderparameter coupling}

The GL potential which include the strain-order parameter
coupling is\cite{Sigrist-Ueda},
\begin{eqnarray}
F_{\rm GL}&=&F_{\rm op} + F_{\rm st-op}+ F_{\rm el},
\nonumber\\
F_{\rm st-op}&=&
-|\eta|^2 C_{\rm A_1} (\epsilon_{xx} + \epsilon_{yy}+ \epsilon_{zz})
\nonumber\\
&&-|\eta_1|^2
\left[C_{\rm A_1} (\epsilon_{xx} + \epsilon_{yy} +
\epsilon_{zz})\right.
\nonumber\\
&&\left. ~~~~~~~~~
+ C_{\rm E} (2 \epsilon_{yy} - \epsilon_{zz} - \epsilon_{xx} )\right]
\nonumber\\
&&-|\eta_2|^2 \left[C_{\rm A_1} (\epsilon_{xx} + \epsilon_{yy} +
\epsilon_{zz})\right.
\nonumber\\
&&\left. ~~~~~~~~~
- C_{\rm E} (2 \epsilon_{yy} - \epsilon_{zz} - \epsilon_{xx} )\right]
\nonumber\\
&&-\sqrt{3}C_{\rm E} (\eta_1 \eta_2^* + c.c.)
(\epsilon_{zz} - \epsilon_{xx})
\nonumber\\
&&-C_{\rm AE} (\eta \eta_1^* + c.c.)(2 \epsilon_{yy} -\epsilon_{zz} - 
\epsilon_{xx}) 
\nonumber\\
&& - \sqrt{3} C_{\rm AE} (\eta \eta_2^* + c.c.)(\epsilon_{zz} 
- \epsilon_{xx}), 
\nonumber\\
F_{\rm el}&=&\frac{1}{2} \sum_{ijkl} c^{(0)}_{ijkl} \epsilon_{ij}\epsilon_{kl},
\label{st-op-coup}
\end{eqnarray}
where $c^{(0)}_{ijkl}$ is the background elastic constants.
Obviously, it is symmetric under the T$_h$ symmetry.

Let us consider the amplitude of the strain. 
It is obtained by solving the equation
$$
0=\frac{\partial F_{GL}}{\partial \epsilon_{ij}}, 
$$
We introduce the notation $
\epsilon_{xx}=\epsilon_1,\epsilon_{yy}=\epsilon_2,\epsilon_{zz}=\epsilon_3,
\epsilon_{xy}=\epsilon_4,\epsilon_{yz}=\epsilon_5,\epsilon_{zx}=\epsilon_6. 
$
In the A-phase (see, Eq. (\ref{A})), 
one finds the isotropic result
\begin{eqnarray}
\epsilon_1&=&\epsilon_2=\epsilon_3=
-\frac{C_{\rm A_1}}{\lambda}\frac{A_{\rm A_1}(T)}{2 \beta}, 
\nonumber\\
\epsilon_4&=&\epsilon_5=\epsilon_6=0,
\label{st-A}
\end{eqnarray}
where $\lambda=c^{(0)}_{11}=c^{(0)}_{22}=c^{(0)}_{33}$.
In B-phase (See, Eq. (\ref{B})), 
an anisotropy can be seen in the $y$ direction:
\begin{eqnarray}
\epsilon_3&=&\epsilon_1=-\frac{C_{\rm A_1}}{\lambda}
\frac{A_{\rm A_1}(T)}{2 \beta} -
\frac{C_{\rm A_1} + C_E}{\lambda} \frac{A_{\rm E}(T)}{2 \beta_1}, 
\nonumber\\
\epsilon_2&=&-\frac{C_{\rm A_1}}{\lambda} \frac{A_{\rm A_1}(T)}{2 \beta} -
\frac{C_{\rm A_1} - 2 C_E}{\lambda} \frac{A_{\rm E}(T)}{2 \beta_1}, 
\nonumber\\
\epsilon_4&=&\epsilon_5=\epsilon_6=0.
\label{st-B}
\end{eqnarray}
It should be noted that the terms proportional to $C_{\rm AE}$ in 
Eq. (\ref{st-op-coup}) play no roles in the state Eq. (\ref{B}).

\subsection{thermal expansion}

The presence of the strain can be measured by the
thermal expansion which is definition by 
\begin{equation}
\alpha_{\mu}=\alpha^{(0)}_{\mu} + \frac{\partial \epsilon_{\mu}}{\partial T},
\end{equation}
where $\mu=1, \cdot\cdot\cdot, 6$, and
$\alpha^{(0)}_{\mu}$ is the background value.
Define $\Delta \alpha_{\mu} = \alpha_{\mu} - \alpha^{(0)}_{\mu}$.
Then, in the A-phase (See, Eq. (\ref{st-A})),
\begin{eqnarray}
\Delta_{\rm A} \alpha_1 &=&\Delta_{\rm A} \alpha_2
= \Delta_{\rm A}\alpha_3 =
-\frac{C_{\rm A_1}}{\lambda}\frac{1}{2 \beta T_{c1}}
\nonumber\\
\Delta_{\rm A} \alpha_4 &=&\Delta_{\rm A} \alpha_5
=\Delta_{\rm A} \alpha_6=0.
\end{eqnarray}
In B-phase (See, Eq. (\ref{st-B})), the thermal expansion becomes anisotropic: 
\begin{eqnarray}
\Delta_{\rm B} \alpha_3&=&
\Delta_{\rm B}\alpha_1=- \frac{C_{\rm A_1}}{\lambda}\frac{1}{2 \beta
T_{c1}} - \frac{C_{\rm A_1} + C_{\rm E}}{\lambda}\frac{1}{2 \beta_1 T_{c2}},
\nonumber\\
\Delta_{\rm B} \alpha_2&=&- \frac{C_{\rm A_1}}{\lambda}\frac{1}{2 \beta
T_{c1}} - \frac{C_{\rm A_1} - 2 C_{\rm E}}{\lambda}\frac{1}{2 \beta_1 T_{c2}},
\nonumber\\
\Delta_{\rm B} \alpha_4 &=&\Delta_{\rm B} \alpha_5 =\Delta_{\rm B} \alpha_6=0.
\end{eqnarray}

The thermal expansion along [100] direction has been
measured\cite{Niels}, and two clear jumps are observed at $T_{c1}$ and
$T_{c2}$. The ratio of these two jumps is
\begin{equation}
\frac{\Delta_{\rm B} \alpha_1 - \Delta_{\rm A} \alpha_1 }
{\Delta_{\rm A} \alpha_1}=\left(1 + \frac{C_{\rm E}}{C_{\rm A_1}}\right)
\frac{\beta}{\beta_1}\frac{T_{c1}}{T_{c2}}\simeq 1.8.
\end{equation}
From Eqs. (\ref{Tc}) and (\ref{b/b1}), one can obtain a relation
\begin{equation}
C_{\rm E} / C_{\rm A_1} \simeq 0.9,
\label{C}
\end{equation}
and estimate the ratio of jumps at the two transition points
for the thermal expansion in the [010] direction 
\begin{equation}
\frac{\Delta_{\rm B} \alpha_2 - \Delta_{\rm A} \alpha_2 }
{\Delta_{\rm A} \alpha_2}=\left(1 -2 \frac{C_{\rm E}}{C_{\rm A_1}}\right)
\frac{\beta}{\beta_1}\frac{T_{c1}}{T_{c2}}\simeq -0.8.
\end{equation}
This quantity has not been measured yet. 

\subsection{isothermal elastic constants}

The isothermal elastic constant is
\begin{equation}
c_{\mu\nu}=
\left(\frac{\partial^2 F_{GL}}{\partial \epsilon_{\mu}
\partial \epsilon_{\nu}}\right)_T.
\end{equation}
The small variation of the strain changes
the amplitude of the order parameters as
\begin{eqnarray}
\delta |\eta|^2 &=&
\frac{C_{\rm A_1}}{2 \beta}(\delta \epsilon_1 + \delta \epsilon_2 + \delta
\epsilon_3),
\nonumber\\
\delta|\eta_2|^2&=&
\frac{C_{\rm A_1}}{2 \beta_1}(\delta \epsilon_1 + \delta \epsilon_2
+ \delta \epsilon_3)
\nonumber\\
&&- \frac{C_{\rm E}}{2 \beta_1}(2 \delta \epsilon_2 -
\delta \epsilon_3 - \delta \epsilon_1 ).
\end{eqnarray}
By using the condition Eq. (\ref{stat}),
\begin{equation}
c_{\mu\nu}=c_{\mu\nu}^{(0)} + \left[
\frac{\partial \vec{\eta}}{\partial \epsilon_{\mu}} \cdot 
\frac{\partial^2 F_{\rm GL}}{\partial \vec{\eta} \partial \epsilon_{\nu}}
 + c.c. \right],
\end{equation}
where $c_{\mu\nu}^{(0)}$ is the background elastic constant.
Let us define $\Delta c_{\mu\nu}= c_{\mu\nu} - c_{\mu\nu}^{(0)}$.
In A-phase,
\begin{eqnarray}
\Delta_{\rm A} c_{11}=\Delta_{\rm A} c_{22}=
\cdot\cdot\cdot=\Delta_{\rm A} c_{31}=-\frac{C_{\rm A}^2}{2 \beta},
\end{eqnarray}
and the others are zero. In B-phase,
\begin{eqnarray}
\Delta_{\rm B} c_{33}&=&\Delta_{\rm B} c_{11}
=\Delta_{\rm B} c_{31}
=-\frac{C_{\rm A_1}^2}{2 \beta}-\frac{(C_{\rm A_1} + C_{\rm E})^2}{2 \beta_1},
\nonumber\\
\Delta_{\rm B} c_{22}&=&-\frac{C_{\rm A_1}^2}{2 \beta}
- \frac{(C_{\rm A_1} - 2 C_{\rm E})^2}{2 \beta_1},
\\
\Delta_{\rm B} c_{23}&=&\Delta_{\rm B} c_{12}
=-\frac{C_{\rm A_1}^2}{2 \beta}
-\frac{(C_{\rm A_1} + C_{\rm E})(C_{\rm A_1} - 2 C_{\rm E})}{2 \beta_1},
\nonumber
\end{eqnarray}
and the others are zero.
From Eqs. (\ref{b/b1}) and (\ref{C}), one can estimate
the ratios of jumps at two transition points of the elastic constants,
\begin{eqnarray}
\frac{\Delta_{\rm B} c_{11} - \Delta_{\rm A} c_{11}}
{\Delta_{\rm A} c_{11}}
&=&\frac{\beta}{\beta_1}\left(1 + \frac{C_{\rm E}}{C_{\rm A_1}}\right)^2
\simeq 3.6,
\nonumber\\
\frac{\Delta_{\rm B} c_{22} - \Delta_{\rm A} c_{22}}{\Delta_{\rm A} c_{22}}
&=&\frac{\beta}{\beta_1}\left(1 -2 \frac{C_{\rm E}}{C_{\rm A_1}}\right)^2
\simeq 0.6,
\nonumber\\
\frac{\Delta_{\rm B} c_{12} - \Delta_{\rm A} c_{12}}{\Delta_{\rm A} c_{12}}
&=&\frac{\beta}{\beta_1}\left(1 + \frac{C_{\rm E}}{C_{\rm A_1}}\right)
\left(1 -2 \frac{C_{\rm E}}{C_{\rm A_1}}\right).
\nonumber\\
&\simeq& -1.4.
\end{eqnarray}

In conclusion for this section, 
the thermal expansion and the elastic constant are 
isotropic in the A-phase, but anisotropic in the B-phase. 
To observe the anisotropy would serve as additional 
evidences for the phase transition and 
the spontaneous cubic symmetry breaking in the B-phase.

\section{Summary and discussions}

In summary, the multiple superconducting phases (A-phase and B-phase)
in the skutterudite PrOs$_4$Sb$_{12}$ with cubic crystal symmetry T$_h$
is discussed by using the phenomenological Ginzburg-Landau approach.
The spin singlet pairing case is considered here.
According to the thermal conductivity measurement\cite{Izawa}
, the gap function has within in the $ab$-plane four point-node-like
structure (four dips) along the [100] and [010]
directions on the Fermi surface in the A-phase, and two dips along the [010]
direction in the B-phase. But the gap structure along the $c$-axis was
unresolved. This problem is considered here.

Because of the absence of an anisotropy in H$_{c2}$\cite{Izawa},
the superconductivity in A-phase appears to be the conventional state
(anisotropic $s$-wave state).
Then, the gap structure has the cubic symmetry 
(the A$_1$ representation) and leads to two additional 
dips along the $c$-axis.
In the B-phase, a state
without additional dips along the $c$-axis
(anisotropic $s + i d_{z^2 - x^2}$-wave state)
is stabilized by the Ginzburg-Landau potential 
of the A$_1$$\oplus$E combined representation
in a wide parameter region. On the other hand, 
it is hard to see that a state with two dips along 
$c$-axis is stabilized, since to obtain 
such a state requires a  
highly accidental mixing between the basis functions 
of the crystal symmetry T$_h$.    
Moreover, it has been shown 
that the $s+ id$-wave state is energetically favored in the A$_1\oplus$E 
combined representation based on the consideration for 
the condensation energy of each state 
within the weak coupling approach\cite{Sigrist-T-broken}. 
Hence it is 
natural to expect that the gap function in A-phase has
a total of six dips along [100],[010] and
[001] directions and while in the B-phase there are in total
two dips along the [010] direction.
The phase transition can be described by the Ginzburg-Landau potential. 

The gap function inthe  B-phase breaks
the cubic crystal symmetry T$_h$ spontaneously. 
This symmetry breaking would cause anisotropic crystal strain 
and anisotropic anomaly of the thermal expansion
and the isothermal elastic constant could be observed . 

{\it The gap function in B-phase 
breaks the time reversal symmetry} (See, Eqs. (\ref{A+E}) and (\ref{2P})).
Then, the spontaneous magnetization around impurities
is expected to be observed by 
the $\mu$SR measurement\cite{Sigrist-Ueda}.  
The magnetization have not been detected yet.
More accurate measurements are highly desired\cite{Maclaughlin}.

Another expected phenomenon in the B-phase
is anomalous flux flow which comes from the domain
formation\cite{Sigrist-Agterberg},
because the domain structure would exist in the B-phase since
the stable state Eq. (\ref{2P}) is two-fold degenerate.

Additional phase transitions induced by uniaxial pressure
may be possible in this superconductor.
Such a transition could occur in a superconductor described by
multicomponent order parameter\cite{Sigrist-Ueda}.

\acknowledgements
The author is grateful to N. Hayashi, K. Izawa, K. Maki,
Y. Matsuda, Y. Nakajima, N. Oeschler, M. Sigrist, A. Tanaka, 
K. Ueda, I. Vekhter,
Q. Yuan, and especially to
P. Thalmeier for useful discussions and comments.

\end{multicols}

\end{document}